\documentstyle[sprocl]{article}

\def\bra#1{{\langle #1 |}}
\def\ket#1{{| #1 \rangle}}

\def\proj{\hat {\cal P}}
\def\id{{\hat 1}}
\def\tr{{\rm Tr}}
\def\H{{\hat H}}
\def\C{{\hat C}}
\def\A{{\hat A}}
\def\B{{\hat B}}
\def\Q{{\hat Q}}
\def\U{{\hat U}}

\begin{document}

\title{Probability in decoherent histories}

\author{Todd A. Brun}

\address{Institute for Advanced Study, Einstein Drive, \\
Princeton, NJ  08540, USA \\
E-mail:  tbrun@ias.edu}


\maketitle
\abstracts{  The decoherent (consistent) histories formalism has been
proposed as a means of eliminating measurements as a fundamental concept
in quantum mechanics.  In this formalism, probabilities can be assigned to
any description which satisfies a particular consistency condition.  The
formalism, however, admits incompatible descriptions which cannot be
combined, unlike classical physics.  This seems to leave an ambiguity in
the choice of the description.  I argue that this ambiguity is removed by
considering the observer as a physical system.}

\section{Introduction}

The most important problems in the interpretation of quantum
mechanics---possibly the {\it only} important problem---is the so-called
measurement problem:  the inconsistency between the unitary evolution
described by the Schr\"odinger equation and the discontinuous, non-unitary
evolution given by the von Neumann projection postulate.  In standard
QM, the wavefunction of a quantum system evolves continuously
according to the Schr\"odinger equation,
\begin{equation}
{d\ket\psi\over dt} = - {i\over\hbar} \H\ket\psi ,
\end{equation}
where $\H$ is a Hamiltonian operator and $\ket\psi$ the state of the
system at the current time $t$.  When the system is measured, however,
by an external measuring device, the state jumps instantaneously
to a new state
\begin{equation}
\ket\psi \rightarrow \ket{\psi_k} = \proj_k \ket\psi/\sqrt{p_k}
\end{equation}
with probability $p_k = \bra\psi\proj_k\ket\psi$, where the $\{\proj_k\}$ are
a complete set of orthogonal projection operators.  The choice of
projections depends on the quantity being measured.

It is clear that these two evolution laws are quite different.  Presumably
the difference arises because of the influence of the external measuring
device.  But any such device must itself be made of atoms and other
components which are themselves subject to quantum laws.  If we include
the measuring device together with the system as a larger, joint system,
this larger system will no longer obey the von Neumann projection rule;
instead, it will evolve into a superposition of all possible measurement
results, reminiscent of the famous Schr\"odinger's cat paradox.

We can attempt to go beyond this by having an observer look at the
measuring device; this itself might count as a measurement, which will
``collapse'' the wavefunction.  But the observer, too, is composed of atoms
and molecules which obey quantum laws.  We seem to be caught in an infinite
regress.

One proposal to get out of this regress is the ``decoherence'' program of
Zurek, Joos and Zeh, and others\cite{Decoherence}.
Any macroscopic system (like a measuring
device) must interact with many microscopic degrees of freedom in its
environment:  stray photons and molecules of gas which bounce off of it,
the atoms in the floor beneath, etc.  These extra degrees of freedom
become correlated with the state of the macroscopic system, destroying
the possibility of macroscopically distinct states interfering with
each other.  In essence, they continuously perform ``measurements'' on
all macroscopic systems with which they interact.

This is an important insight, and is unarguably a real effect:  the
experimental evidence for decoherence is overwhelming.  But as a solution
for the measurement problem it leaves many people dissatisfied.  If we
describe the state of the system alone, tracing out the environment,
decoherence can indeed explain how an initially pure state $\ket\psi$
can evolve into a mixed state
\begin{equation}
\rho = \sum_k \ket{\psi_k} p_k \bra{\psi_k}
  = \sum_k \proj_k \ket\psi\bra\psi \proj_k ,
\end{equation}
which looks like a probabilistic mixture of different measurement outcomes
$\ket{\psi_k}$ with probabilities $p_k$, just as in the measurement
scheme described above.  But critics complain that there is still an
unexplained step between getting such a density matrix $\rho$ and getting
a {\it single} outcome $\ket{\psi_k}$.

Moreover, these same critics complain that even getting this density
matrix depends crucially on making the subjective distinction between system
and environment, and on tracing out the environment degrees of freedom.
If we describe the state of system and environment together, it remains
pure, and obeys Schr\"odinger's equation at all times.  What is missing
is an explanation of our subjective experience, in which a single event
occurs with some probability.

In order to explain this experience, we need to have some idea of what a
probability is.  Common explanations in terms of the frequencies of repeated
events are unsatisfactory; most events are {\it not} repeated exactly,
and even if they are, for a finite number of repetitions it is always
possible that the observed frequencies will be very different from those
that would be predicted {\it a priori}.

A more satisfactory definition is that of the {\it Bayesians}:  the
probability of an outcome is a measure of our certainty as to whether that
outcome will occur.  This sounds quite subjective, but in a sense it is not.
Given the same prior information, any two rational beings should assign
the same probabilities to the same outcomes\cite{Jaynes}.

By combining this notion of subjective probabilities with a formulation
of quantum mechanics which enables a discussion of whether or not events
occur, it is possible to form an internally consistent description
of quantum mechanics without any special measurement postulate.  This
formulation explains both the usual freedom of quantum mechanics to describe
systems in terms of any choice of observables, including macroscopic
superpositions, and our subjective experience in which only a single
macroscopic state occurs.  In the following notes I develop this argument
using the consistent histories formalism of quantum
mechanics\cite{Griffiths,Omnes,GellMann}.
I think any argument along these lines will arrive at a similar conclusion
(including the fact that the histories which describe our possible
experiences form a consistent set).  However, I do not claim that this is
the only way in which this formalism may be consistently interpreted.

The decoherent histories formalism has been attacked as `ambiguous' because
it admits multiple incompatible descriptions of the same quantum
system\cite{DowkerKent}.
I will argue in this paper that this criticism is misguided, and based
on a confusion between descriptions and the things they describe.  We
must immediately understand two important points.  First, the `incompatibility'
between different descriptions in no way implies that they {\it contradict}
each other, but is a technical term indicating that they cannot be combined
into a single, more fine-grained description.  This is unintuitive---in
classical physics, such a combination is always possible.  But it is not
inherently paradoxical.

Second, this ambiguity of description is a freedom we enjoy as theorists;
but it does not imply any ambiguity in the answer of unambiguous questions.
Given a particular physical system and a particular question about it,
any consistent description which addresses this question will given
exactly the same answer.  In particular, we human beings are physical
systems before we are theorists; and while we may entertain many possible
descriptions of the world in our minds, we have no choice about what we
actually experience.

Because I don't wish to consider the philosophical problem of explaining
our conscious experience, which may indeed be beyond the scope of physics,
the protagonist in these notes will be a robot, equipped with a computer
brain and memory and detectors which serve as its senses.  This is
similar to Jaynes's\cite{Jaynes} use of a robot to emphasize that
any rational being will assign the same probabilities given the same prior
information.

Finally, let me clarify that `observers' are in no way necessary for
the understanding of quantum mechanics.  I treat the robot in this paper
solely as a model for understanding how, in principle, we might use
quantum mechanics to unambiguously predict our own subjective experience.
In practice, we usually use quantum mechanics to describe systems
without observers.

\section{Consistent histories and branching wavefunctions}

The formalism of consistent histories is well known, so we reprise
it only briefly here.  Suppose that a closed quantum system
is initially in a state $\ket{\psi_0}$.  It is
then possible to choose a succession of times $t_1 < t_2 < \ldots < t_N$,
and at each time specify an exhaustive set of alternatives at each time $t_i$,
represented mathematically by a set of orthogonal projections
$\proj^i_{\alpha_i}$ which give a decomposition of the identity:
\begin{equation}
\sum_{\alpha_i} \proj^i_{\alpha_i} = \id,\ \
\proj^i_{\alpha_i} \proj^i_{\alpha_i'} = \delta_{\alpha_i \alpha_i'}.
\end{equation}
A {\it history} then consists of an alternative at each time.
The history operator is defined
\begin{equation}
\C_\alpha \equiv \proj^N_{\alpha_N}(t_N) \cdots \proj^1_{\alpha_1}(t_1),
\end{equation}
where $\proj^i_{\alpha_i}(t_i)$ is the Heisenberg operator
$\exp[i\H t_i] \proj^i_{\alpha_i} \exp[-i\H t_i]$.

A set of histories is {\it consistent} if it satisfies the criterion
\begin{equation}
D[\alpha,\alpha'] = \tr\{ \C_\alpha \ket{\psi_0}\bra{\psi_0}
  \C^\dagger_{\alpha'} \} = \delta_{\alpha\alpha'} p(\alpha),
\label{consistency}
\end{equation}
for all pairs of histories $\alpha$ and $\alpha'$.  If this is satisfied,
then the diagonal terms $p(\alpha)$ can be interpreted as the
{\it probabilities} of the histories $\alpha$, and these probabilities
satisfy the usual probability sum rule.

Such a set of histories forms a branching structure.  At time
$t_0$ we know only the initial state of the universe; at time $t_1$
we split this into a number of different alternatives; each of these
is in turn split at time $t_2$, and so forth.  Much has been made of this
branching process, with some arguing that a physical mechanism must exist
to select one branch and discard the others.  Unfortunately, there is no
single, unique consistent set of histories.  Any set which
obeys the consistency criterion (\ref{consistency}) is as valid a choice
as any other.  The theory is silent on this point, a fact which has
sometimes been cited as a fatal flaw.

Actually, there is a simple way of understanding this multiplicity
of consistent sets in terms familiar from standard quantum mechanics,
by appreciating that specifying a consistent
set of histories is the same as {\it resolving the evolving wavefunction
into orthogonal components at all times.}  This resolution into components
has the same branching structure as the histories:  if there are
$n_1$ alternatives at time $t_1$, $n_2$ at $t_2$ and so forth, then
at times $t<t_1$ there is only one component; at $t_1 < t < t_2$ there
are $n_1$ components; at $t_2 < t < t_3$ there are $n_1 n_2$ components,
and so forth.  The consistency criterion ensures that the components are,
in fact, orthogonal.

If $t$ is between $t_j$ and $t_{j+1}$ then the wavefunction can be
written
\begin{eqnarray}
\ket{\psi(t)} = && \exp[-i\H t] \ket{\psi_0} \nonumber\\
= && \sum_{\alpha_1,\ldots,\alpha_j} \exp[-i\H(t-t_j)] \proj^j_{\alpha_j}
  \exp[-i\H(t_j-t_{j-1})] \proj^{j-1}_{\alpha_{j-1}} \cdots
  \proj^1_{\alpha_1} \nonumber\\
&& \times \exp[-i\H t_1] \ket{\psi_0}.
\end{eqnarray}

Because Hamiltonian evolution preserves orthogonality, if we have chosen
a particular resolution of the wavefunction into orthogonal components
we can (if we like) select a single component, renormalize it, and follow
its evolution without having to worry about any of the others.  It is
this fact which enforces obedience to the probability sum rules.  Since
any set of orthogonal components can, in standard quantum mechanics,
be considered eigenstates of an observable, an appropriate series of
measurements would pick out exactly one final component, with a
probability equal to the probability of the history.  But in consistent
histories it is unnecessary (and meaningless) to invoke a measuring
device outside the system.

Thus, we see that the sort of intuitive picture often invoked in
discussions of `Many-Worlds,' in which the wavefunction repeatedly
branches, makes sense when considered in the context of consistent
histories.  It is not clear, however, that it makes sense to talk of
these branches all being real; the reality of something with which one
can never interact seems more a question for philosophy than physics.

There is one major caveat here.  There is by no means only a
single way of resolving the wavefunction into orthogonal components.
Indeed, there is an infinite number of ways, corresponding to the
infinitude of consistent sets.  On the level of orthogonal components,
this is little more than the statement that many bases can be chosen
for each branch.

This multiplicity of descriptions implies no physical inconsistency;
we are, in a sense, visualizing the universe from the outside, and
can do so in any way that we choose.  However, that is not to say
that there is no physical significance attached to particular
descriptions, or choices of set.  If one wishes to discuss, for example,
the value of a particular physical variable, it only makes sense to do
so in the context of histories which assign it a value.  This should
be completely obvious, but has given rise to great confusion.

Thus, there seem to be two ways of looking at consistent histories.
From the outside, it seems to be a decomposition of a unitarily
evolving wavefunction into numerous coexisting components.  But in
the context of a single consistent set, one might equally well think
of it as a stochastic model in which one history occurs with a given
probability, and the others represent only potential outcomes.
I will argue in this paper that it is impossible for an observer
inside a closed `universe' to distinguish between these two
pictures, provided only that we assume the weights assigned to
single histories correspond to subjective probabilities of the
observer.

\section{Bayesian probabilities}

From the preceding discussion it is clear that probabilities, whatever
they are, must arise at the level of the branching.  However, given that
there are many different consistent sets, there are many ways in which
this branching could be considered to occur.  What does it mean, then,
to assign probabilities to these branches?

A good step towards answering this question is to ask what it means
to assign probabilities to alternatives in {\it classical} physics.
Surprisingly, there is still considerable controversy on this point.
For years there has been an ongoing debate between the {\it frequentist}
and {\it Bayesian} interpretations of probability theory.  I sketch
them both, briefly.

In the frequentist picture, the probability of a given result is the
frequency with which that result occurs over a large number of repeated
trials.  This has a certain intuitive appeal, and lends itself well to
describing some problems, such as the odds of rolling various numbers
with dice.

Unfortunately, as a rigorous basis for probabilities the frequentist
description has serious flaws.  What does it mean to say that the
probability is the frequency over many trials?  How many trials?  If it
is a finite number, there will always be some cases in which the
frequencies deviate markedly from the underlying probabilities.  How
do we deal with those cases?  One can't simply dismiss them as improbable;
probability is what we are trying to define!

A related problem is that the frequentist approach is mute in assigning
probabilities to single events.  Looking at repeated trials makes sense
when betting on dice, but not when betting on horse-races or football
games; no two races or games will ever be exactly alike.  But that doesn't
stop the bookmakers from setting odds.

The Bayesian interpretation is quite different.  In this picture, a
probability is always {\it subjective}, in the sense that it reflects
the uncertainty of a rational agent with incomplete information.
This agent need not be ``intelligent;'' it need only be able to reason
consistently according to fixed rules.  An appropriately programmed
computer would be a perfectly good rational agent.  Probabilities are
subjective, in that they depend on the information possessed by the
agent, but they are {\it objective} in that any two agents given the
same information would assign exactly the same probabilities, as long
as both agents were rational.

In the Bayesian interpretation all probabilities are a result of
imperfect information, and therefore it makes as much sense
to assign probabilities
to single events as to long sequences.  And what is more, the
correspondence between frequencies and probabilities now becomes
clear:  given the probability for a {\it single} trial, one can deduce the
probabilities that different frequencies will be observed in repeated
trials (as long as these trials are independent).
As the number of trials increases, it becomes less and less
probable that the frequency will deviate significantly from the
single-time probability.

In this way, frequencies remain important.  From the outcome of a single
event, a rational agent has little way of assessing how good its
{\it a priori} probabilities were.  By examining the outcomes of many
events, however, the agent can either gain confidence in its
assessment, or else improve it in light of experience.

This is the Bayesian picture for a classical world.  It has had great
success in unifying the results of probability theory within a single,
consistent framework.  But how must it be changed to deal with a
fundamentally quantum world?

\section{The robot}

The first question to be answered is what exactly is a rational
agent?  A rational agent is simply a physical system which is
capable of processing information according to definite rules:
in short, a computer.  Following E.T. Jaynes, we term this agent
`The Robot'\cite{Jaynes}.

Since we imagine our robot to be something which could (at least
in principle) exist, we model it as a {\it finite automaton}.  Such a
device has a finite number of possible internal states, and a
finite number of possible inputs.  At each stage in its computation
it receives a single input value, and based on its current state and the
value of the input it evolves deterministically to a new state.

The robot begins with a certain amount of prior information.  This
is contained both in its programming (i.e., the rules by which it
changes states) and its initial state.  We assume that it has been
programmed to reason consistently, and is therefore rational in our
limited sense.

The internal state of the robot is a valid observable, so we can
choose basis states which are eigenstates of this
observable.  We label these basis states $\ket{B_n}$, where
$\B$ is the observable and $B_n$ indicates that the robot is in
its $n$th internal state.

This observable $\B$ is highly coarse-grained.  The robot will
undoubtedly contain many more internal degrees of freedom which
are more or less irrelevant for our purposes.  We will call these
degrees of freedom the `environment of the robot' and label them
$b$.  Thus a complete state of the robot could be expressed in
a basis $\ket{B_n,b}$.

This is not sufficient to describe the functioning of the robot.
It must also receive data from the outside world.  This data will,
in general, be far from a complete description of the world.  Rather,
we assume that the robot's senses are limited, so that it can only
get a very coarse-grained picture.  Let $\A$ be the observable
which the robot has access to.  For instance, $\A$ might be the output
from a measuring device, or a group of measuring devices.  We
call any other degrees of freedom which are irrelevant to the value of $\A$
`the environment of $A$,' and label them $a$.  Thus, a state of the
robot, its input data, and their respective environments
can be expressed in the basis
$\ket{A_m,a,B_n,b}$.

Finally, there may be other degrees of freedom in the universe to
which the robot has no direct access, but which {\it do} affect the dynamics
of $\A$.  We will lump these together under the label $\Q$,
with eigenvalues $Q_l$.  Note that there may be more than one reasonable
choice of $\Q$; for example, if the external system were a spin-1/2
particle one might choose to express the spin in the $x$ basis, the
$y$ basis, or any other direction.  Picking one description $\Q$
for the present, a complete
state can now be written
\begin{equation}
\ket\psi  = \sum_{l,m,n} c_{l,m,n} \ket{A_m,a,B_n,b,Q_l}.
\end{equation}
We will usually suppress the environment labels $a,b$.  They can be
important, however, in that they allow the dynamics of $A$ and $B$ to
be irreversible and decoherent.

In describing the dynamics of the robot and its world, we let time
be discrete, each time corresponding to a single tick of the robot's
internal clock.  At each step, the robot will change from its current
state to one of $N$ possible successor states, depending on which if
its $N$ possible values the observable $\A$ assumes.  The state of
the robot after $j$ steps then depends on its initial value and the
succession of values $A_i$ that it observes:
\begin{equation}
B(t_j) = B(B(t_{j-1}),A_{j-1})
  = B(B_0, A_0, A_1, \ldots, A_{j-1}).
\end{equation}
We assume that the robot starts in a special initial state, ready
to solve a problem.  Note that this dependence on all previous
values of $A_i$ means that the robot ``remembers'' what values of
$\A$ it has already seen.  $\A$ will generally have dynamics of its
own, which for the moment we will assume are independent of $\B$.
(Later we will allow the robot to act on the information
it acquires.)  In a single time step $A_m$ goes to some new
value $A_{m+1}$.

The quantum dynamics is almost exactly the same.  Neglecting, for
the moment, the existence of any significant variables $\Q$,
evolution in time is given by a unitary operator $\U$ which
effects
\begin{eqnarray}
\U\ket\psi = && \U\biggl( \sum_{m,n} c_{m,n} \ket{A_m,B_n} \biggr)
  = \sum_{m,n} c_{m,n} \U \ket{A_m,B_n} \nonumber\\
= && \sum_{m,n} c_{m,n} \ket{A_{m+1},B(B_n,A_m)}.
\end{eqnarray}
To be truly consistent, we should include the environment degrees
of freedom as well:
\begin{equation}
\U\ket{A_m,a,B_n,b} = \ket{A_{m+1},a'(A_m), B(B_n,A_m), b'(B_n)}.
\end{equation}

\section{Histories of the robot}

We are now in a position to ask what our robot `sees' in a given
situation.  The obvious way to do this is to choose for our alternatives
projections onto the internal state of the robot $\proj_{B_n}$.

Note that we need not project onto the internal state of the robot
alone.  We can, if we like, project onto the values of the observable
$\A$ as well, or even include portions of the environments $a$ and $b$
and possible $\Q$ as well, provided that consistency is not violated.
The possible experiences of the robot do not pick out a unique
consistent sets, but rather a large family of such sets, each a
fine-graining of the coarsest description which includes projections
on the internal state of the robot and nothing else.  Some of these
fine-grainings may be incompatible with each other, but they are
all compatible with this coarsest set.
The important point is that histories incompatible with the observable
$\B$ tell us {\it nothing} about what the robot `sees' and `thinks.'

Because $\A$ and $\B$ are perfectly correlated, histories of $\A$, $\B$,
and $\A$ {\it and} $\B$ will all have the same probabilities.  Thus,
it doesn't matter whether we take the worm's-eye-view inside the robot's
brain, or the `objective' picture of what it considers is going on
in the world outside.  The predictions in either case are the same.

This, by the way, is a good place to point out that this robot is
{\it not} intended to fill a role similar to that filled by `The
Observer' in standard quantum mechanics.  There is no mystical
significance to the presence of a rational agent.  It is simply another
physical system, and obeys exactly the same laws as any other physical
system.  I will say more about invoking imaginary `Observers' in a
later section.

\section{Predicting the outcome of an experiment}

Suppose that the robot is programmed to be a gambling machine,
betting on the outcome of a quantum measurement.  The initial state is
\begin{equation}
\ket\psi = \ket{A_0,B_0}\otimes(\alpha\ket{Q_1} + \beta\ket{Q_2}).
\end{equation}
The external system is in a superposition of two eigenstates $Q_1$
and $Q_2$ of some observable $\Q$; for instance, it might be a
spin-1/2 particle.

The robot has the following initial information:  (1) it knows the rules
of quantum mechanics, (2) it knows the initial state of the external
system, and (3) it has been offered some odds $O$ on the outcome of
a measurement of $\Q$.  The variable $\A$ gives the position of the
pointer on a measuring device.

The robot must decide whether or not the odds it has been offered are
fair.  If they are, it will bet \$1 on the outcome $Q_1$; if not, it
will bet nothing.  If the outcome {\it is} $Q_1$, the robot wins
$\$O$.  If it is $Q_2$, it loses a dollar.  The robot's expected winnings
are $O|\alpha|^2 - |\beta|^2$.  Rationally, the robot should only accept
the bet if $|\beta|^2/|\alpha|^2 \le O$.  With this strategy, it can never
lose money on average.

In another way of looking at this, however, it seems like the robot
always both wins {\it and} loses.  After all, if we think of a set
of histories as a branching wavefunction, after the measurement
{\it both} components are present with probability 1.  The evolution
is, in fact, completely deterministic.

This is where it is important to bear in mind that the probabilities
that matter to the robot are {\it subjective} probabilities.  A mythical
`outside observer' might see both outcomes; but there {\it is no such
observer}.  The only observers are inside the system, and an
observer in a given branch can only see the events within that
branch.  The winning and losing robots can never interact or be
aware of each other in any way.

An analogy due to Simon Saunders is helpful in thinking about
this\cite{Saunders}.  (Interestingly enough, almost exactly the same idea
was used by two science fiction authors, Frederick Pohl and Jack Williamson,
in a pair of novels they wrote together\cite{PohlWilliamson}.)
Suppose that we have a (classical) perfect copying machine.  Any object
placed in it is exactly duplicated, without itself being changed in
any way.

Suppose the technicians approach our robot, which we will call A,
and ask it to allow them to duplicate it.  They assure the robot
that it will notice nothing at all.  The robot agrees, enters the
copying machine, and instantaneously a new robot is produced,
called B.  Suppose that B appears on a distant planet, so that
B and A can never compare notes.  Just as the technicians said, A
doesn't feel a thing, and goes on its merry way.

B's experience, however, is quite different.  It, too, remembers being
assured by the technicians that it would feel nothing; but despite
their assurances it now finds itself on a distant planet.  To B,
the device seems less like a copy machine, and more like an
instant transportation device.  If once again asked to be duplicated,
B will undoubtedly think of things quite differently.  Instead of
walking in and walking out unchanged, it will have a 50/50 {\it
subjective} probability of remaining behind or being transported.

The situation of the robot in a quantum universe is quite similar.
The wave function may branch into a superposition of many robots
making different observations, but each one perceives only its
own branch.  Thus, in making a decision before the branch occurs,
the robot should rationally try to maximize the benefits of all
the copies.  This is exactly the same as estimating the subjective
probabilities of each branch.

Only one requirement is necessary for this identification to be
complete:  that the `weights' of the different branches equal the
subjective probabilities of rational agents in those branches.
This could be considered an axiom of consistent histories, but in
fact it may follow directly from the deeper structure of quantum
theory.  Gleason's theorem seems to argue that such an identification
is essentially inevitable.

\section{Estimating the wavefunction---quantum coin tossing}

This situation is parallel to the above case, but differs in important
respects.  Suppose now that the external degrees of freedom consist
of $N$ identical two-level systems in exactly the same initial state,
and that these will be measured successively by the measuring device
with output value $\A$.  This time the robot knows the rules of quantum
mechanics and it knows that all the external systems are in the same
state, but does not know what that state is.  Its task is to estimate
the initial state from the results of $N$ measurements.

Suppose the initial state is
\begin{equation}
\ket{\Psi_0} = \ket{A_0,B_0}\otimes\bigl(\alpha\ket{Q_1}+\beta\ket{Q_2}\bigr)
  \otimes \cdots \otimes\bigl(\alpha\ket{Q_1}+\beta\ket{Q_2}\bigr),
\end{equation}
that is, there are prepared $N$ identical copies of the microscopic
system, each in the same initial state, the output of the measuring
device is in its initial (null) state, and the robot is in the
starting state.  At each subsequent time, the measuring device
measures one of the microscopic systems and produces an output
$\ket{A_1}$ or $\ket{A_2}$; the robot, in turn, observes the output,
and undergoes a transition to a new state.  This state depends on
all of the observations up to that time, and includes an estimate
of the microscopic state with an attached confidence limit.  We can
label the internal state of the robot $\ket{B_{i_1i_2\ldots i_n}}$
after $n$ time steps, with each of the $i_1\ldots i_n$ being
either 1 or 2.  Thus, after $n$ steps the state of the whole system is
\begin{eqnarray}
\ket{\Psi_n} && = \left( \sum_{i_1\ldots i_n} \alpha^{n_1}\beta^{n-n_1}
  \ket{A_{i_n},B_{i_1\ldots i_n}} \otimes
  \ket{Q_{i_1}} \otimes \cdots \otimes \ket{Q_{i_n}} \right) \nonumber\\
&& \otimes\bigl(\alpha\ket{Q_1} + \beta\ket{Q_2}\bigr) \otimes
  \cdots\otimes \bigl( \alpha\ket{Q_1} + \beta\ket{Q_2} \bigr).
\end{eqnarray}
where $n_1$ is the number of $i_1\ldots i_n$ equal to 1, and
$N-n$ of the microscopic systems remain in the initial state.

Clearly projections onto the state of $\A$ at each time form
a consistent set of histories, as do projections on $\B$, or
$\A$ and $\B$.  There will clearly be certain histories in which
the robot gets a very distorted estimate of the initial state.
For instance, in the history where the robot measures $Q_1$ every
time, it will conclude with high confidence that the state is
very close to $|\alpha|=1$.  But if $|\alpha| < 1$, the probability
of this happening is only $|\alpha|^{2N}$.  In most histories, the
robot gets a reasonably accurate picture.

The scheme as described actually only lets the robot estimate the
values of $|\alpha|^2$ and $|\beta|^2$, without their relative
phase.  In a more sophisticated experiment, the robot might have
access to three measuring devices, measuring the $x$, $y$, and
$z$ axes, and would use 1/3 of the prepared systems in each
detector; or might even allocate systems to different detectors based
on its current estimates of its uncertainty.  But the essential situation
is unchanged.

\section{Quantum vs. classical uncertainties}

The two cases described above typify the difference between quantum
and classical probabilities.  In the first case, the robot had total
knowledge of the initial state.  Its uncertainty was completely due
to the inherent indeterminism of quantum systems.  In the second case,
its uncertainties were due to its imperfect information, and hence
were essentially classical in nature.

This illustrates the remarkable characteristic of quantum probabilities:
even {\it maximal} information in incomplete, in the sense that it does
not allow the robot to predict the outcome with certainty.  Consider the
following two experiments.  In the first, each of the $N$ microscopic
systems is prepared in the same state $\alpha\ket{Q_1} + \beta\ket{Q_2}$.
In the second, $|\alpha|^2 N$ of the systems are prepared in state
$\ket{Q_1}$ while $|\beta|^2 N$ of the systems are prepared in state
$\ket{Q_2}$, distributed randomly.  If only allowed to measure $Q_1$
vs. $Q_2$ the robot is unable to distinguish these two situations, but
given the freedom to measure any linear combination of the two it can
quickly tell them apart.  Indeed, if the robot chooses to measure
the $\alpha\ket{Q_1} + \beta\ket{Q_2}$  axis, in the first case it
will always get the same result.  For this particular experiment,
maximal information translates into a deterministic outcome.  But in
general it does not.

Within the context of the lab, the robot has considerable freedom of
choice as to how it will prepare and measure microscopic systems.  But
in the world as a whole it does not, and there is no guarantee that the
variables with which its senses are correlated will correspond to the
exact state of the universe.  Therefore, in general, the robot will
perceive a probabilistic universe, with unavoidable uncertainties.

Given this fact, an interesting question becomes not why is there
so much apparent randomness in the universe, but why is there so
little?  Why do deterministic classical laws hold with such good
precision on the macroscopic level?  The answer to this question
is still not fully understood.  But it seems clear from our
experience that certain variables (highly coarse-grained ones,
for the most part) are much more predictable than others; and
therefore, a well-informed and programmed robot can make much
better judgments about their behavior than it could about general
quantum variables.  It is this fact, indeed, that underlies the
assumption that the robot itself can be described in
quasiclassical terms.  We will briefly examine this question
below.

\section{Discovering quantum theory}

We have, up to this point, assumed that the robot's programming included
a knowledge of the laws of quantum mechanics.  But suppose we wished for
the robot to discover those laws in the first place.  How might it set
about the task?

Of course, characterizing the entire process of scientific research and
discovery is far beyond our abilities; the real discovery of quantum
mechanics was the result of many people working on many different
problems.  But we can consider a `baby' version of this problem.
Suppose, once again, the robot is provided with $N$ microscopic systems,
and the robot must decide between two
physical pictures.   Either the state is a ray in a two-dimensional
Hilbert space $\alpha\ket{Q_1} + \beta\ket{Q_2}$, or it is a classical
spin pointing in a random direction, with definite probabilities
$p_x$, $p_y$, and $p_z$ of the having positive spin components
$S_x$, $S_y$, and $S_z$.  The robot can measure spins along either
the $x$ or $z$ directions, and can perform repeated measurements on the
same spin.

One can sketch out the form of an experiment.  Both pictures give identical
predictions for single spin measurements $x$ or $z$, or pairs of spin
measurements $xx$, $zz$, $xz$ or $zx$.  Their predictions differ, however,
for three measurements $xzx$ or $zxz$.  In the classical picture, the
first and last measurements should always give identical results, while
in the quantum picture they should not.

We now let the robot go, and it performs its $N$ measurements.  In the
vast majority of cases, it will correctly conclude that the quantum
description is better than the classical.  There is a chance, however,
of $1/2^N$ that the experimental outcome will exactly match the
classical prediction.  In that case, the robot will conclude, based on
its available information, that the classical description is much more
likely than the quantum.

In any probabilistic theory there is always a possibility that, purely
by chance, one may reach the wrong conclusion from a correct experiment.
In medical testing, where small samples are the rule, this in fact
happens all the time.  In physics, we rarely spend much time worrying
about the possibility.  After all, the probability of it occurring is
almost unimaginably small for a typical experiment.  But in enumerating
all possible histories, there will always be some, of low probability,
in which highly atypical things happen.  It should be clearly understood
that the existence of such possible histories in no way violates either
the laws of physics or of common sense.

\section{Robots as canonical observers}

I have argued, I hope convincingly, that quantum mechanics (and
decoherent histories in particular) can describe the experiences
of observers without ambiguity.  However, an extreme subjectivist
might now charge that we have undermined the ability of quantum
mechanics to describe a system {\it without} observers.  The formalism
can easily be applied to such systems, to calculate a consistent
set of histories with appropriate probabilities.  But if we
insist that probabilities are meaningful only as the subjective
judgments of a rational agent, what do the probabilities mean if
there are no rational agents around?  Do they mean anything at all?

Probably very few people, if any, would take such an extreme view.
It is the probabilistic equivalent of asserting that a tree falling
in a forest with no one around not only makes no sound, but doesn't
even exist.  However, even without going to such an extreme, one might
still ask for an interpretation of the probabilities we calculate.

One way of answering this question is to invoke an imaginary
`canonical observer,' who passively observes which set of events
occurs, and whose subjective probabilities would match those of
the particular consistent set we are using.  An important property of
consistent sets is that it is always possible, in principle, to add
such an observer to the system without altering the predicted probabilities
of the different histories.  Indeed, it is possible to derive the consistency
criterion from this requirement.  Such an observer would be similar to
the canonical observers, each equipped with a clock and meter stick,
which are often invoked in General Relativity to explain the meaning
of the metric.  These relativistic observers are assumed to be too small
to distort the solution of Einstein's equations.  Similarly, the quantum
observers avoid interfering with the probabilities by restricting
themselves to consistent sets.

The `canonical observers' could be one or more of our robots,
carefully tailored to the particular decoherent set we are using.
But the most important thing to remember is that
these are {\it imaginary} observers.  We do not insist when
doing a calculation in General Relativity that it only makes
sense if space is filled with tiny people carrying clocks and meter
sticks.  Similarly, in quantum mechanics it is perfectly sensible
to create descriptions in which nothing resembling a measurement or
an observer is present.

\section{Why quasiclassical variables?  The parable of the hourglass}

Quasiclassical variables are the familiar variables of the
classical world:  coarse-grained center of mass positions and
momenta of macroscopic objects, averaged field strengths in small
cells in space, and so forth.  These are the variables which most
simply describe us as physical systems, as well as what we
observe in the world around us.

The question is, why should this be so?  Within decoherent
histories there are an infinity of consistent sets, corresponding
to an infinity of possible descriptions, almost none of which are
anything like quasiclassical.  What is it about the
quasiclassical description which is special?  Or is there nothing
special about it at all, and we could have evolved to use a very
different decomposition of the wavefunction?

While the answer to this question is not known, it has been
speculated that what makes the quasiclassical description is its
predictability.  Quasiclassical variables give a highly
coarse-grained description which approximately obeys a closed set
of deterministic equations.  Highly nonclassical descriptions,
such as descriptions in terms of macroscopic superpositions, do
not.

The most famous example of a macroscopic superposition is
Schr\"odinger's cat.  Sealed in a box with a vial of poison,
whose release is controlled by the decay of a single atom, the
cat evolves into an equally-weighted superposition of being alive
and dead:
\begin{equation}
\ket{\psi} = \ket{\rm live,undecayed} \rightarrow \ket{\psi'} =
\frac{1}{\sqrt2}\left(
  \ket{\rm live,undecayed} + \ket{\rm dead,decayed} \right) \;.
\end{equation}
If we describe this system in the decoherence formalism, we could
choose a set which includes projectors $\ket{\psi}\bra{\psi}$ and
$\ket{\psi'}\bra{\psi'}$ at the initial and final times.  Such a
description is highly nonclassical, but obviously consistent
(since $\ket{\psi'}$ is just the unitarily evolved successor to
$\ket{\psi}$).  Or we could choose a quasiclassical description,
with projectors $\proj_{\rm live},\proj_{\rm dead}$ at both the
initial and final times.  Why should we choose one rather than
the other?

One observation which we should make is that live vs. dead is a
very coarse-grained trait.  The projectors $\proj_{\rm live}$ and
$\proj_{\rm dead}$ correspond to very large subspaces of the
Hilbert space of the cat.  The quasiclassical description is thus
robust under perturbations of the initial state, the dynamics,
and the times of the projections. The unitary description is not.

The time evolution of the quasiclassical description is also far
simpler.  In the case of Schr\"odinger's cat, it begins by being
alive; if we wait long enough (and the humane society doesn't
intervene) the cat will become dead, and remain dead thereafter.
The description in terms of macroscopic superpositions, by
contrast, will change constantly and rapidly, exhibiting extremely
complicated dynamics.

The physics of live and dead cats is a bit too complicated for
easy analysis, but the essential point can be captured by a
classical analogy.  Consider an hourglass, which begins with all
the sand in the upper half.  After approximately an hour, all the
sand will have dropped to the lower half.  We could try to
describe this system by keeping track of the position, velocity
and orientation of every grain of sand, but this is far too
complicated to carry out in practice.  Instead, we might consider
some kind of coarse-grained description of the hourglass.

Here are two such coarse-grainings, which are superficially
similar.
\begin{eqnarray}
f(t) &=& \cases{1&if more sand on top at time $t$;\cr
  0&otherwise.\cr} \nonumber\\
g(t) &=& \cases{1&if odd number of grains on top at time $t$;\cr
  0&otherwise.\cr} \;.
\end{eqnarray}
Both variables are defined at all times, and give exactly one bit
of information about the state of the hourglass.  But $f(t)$
gives a simple description with a simple time-evolution, which is
robust under perturbations of the initial state; the exact time
of the transition from $1$ to $0$ may vary slightly, but the
essentials of the description are unchanged.  By contrast, $g(t)$
exhibits very complicated behavior, which evolves unpredictably
on a much shorter timescale than $f(t)$, and which is highly
sensitive to the exact initial state of the sand.  Which
description is simpler and more stable?  Which is more useful? It
is not hard to see that quasiclassical histories are more like
$f(t)$, and histories of macroscopic superpositions are more like
$g(t)$.

\section*{Acknowledgments}

I would like to acknowledge discussions over many years with
Murray Gell-Mann, Bob Griffiths, Jonathan Halliwell and Jim
Hartle, which have very materially shaped my ideas on this subject
(though the viewpoint given here is no one's fault but my own). I
have also been influenced by conversations with Robert Garisto,
Seth Lloyd, Simon Saunders and R\"udiger Schack, for which I am
very grateful.  My research in this area has been supported by
the Martin A.~and Helen Chooljian Membership in Natural Sciences,
and by DOE Grant No.~DE-FG02-90ER40542.


\end{document}